\documentclass[aps,prev,twocolumn,preprintnumbers,floatfix,nofootinbib]{revtex4-1}
\usepackage[english]{babel}
\usepackage{bm}
\usepackage{times}
\usepackage[normalem]{ulem}
\usepackage{listings}
\usepackage{slashed}
\usepackage{color}
\usepackage{appendix}
\usepackage{mathtools}
\usepackage{natbib}
\usepackage{epsfig}
\usepackage{dcolumn}
\usepackage{textcomp}
\usepackage{multirow}
\usepackage{graphicx}
\usepackage{soul}
\usepackage{subfigure}
\usepackage{xcolor}
\usepackage{comment}
\usepackage{amsmath}
\usepackage{hyperref}  
\usepackage{booktabs}
\usepackage{float}
\usepackage{amssymb}
\usepackage{slashed}
\usepackage{subfigure}
\usepackage[compat=1.1.0]{tikz-feynman}
\usepackage{tabularx}
\usepackage{hyperref}

\allowdisplaybreaks

\begin{document}
\title{Cosmological Constraints on Long-Lived Particles Using Dimension-Six Effective Operators}

\author{Mickael V. S. de Farias$^{1,2}$}
\email{}
\author{Rodrigo Holanda$^{1}$}
\email{}
\author{Matheus M. A. Paixão$^{1,2}$}
\email{matheus.mapaixao@gmail.com}
\author{Farinaldo S. Queiroz$^{1,2,3,4}$}
\email{farinaldo.queiroz@ufrn.br}
\author{Priscila V. dos Santos$^{2,5}$}
\email{}

\affiliation{$^{1}$Departamento de Física, Universidade Federal do Rio Grande do Norte, 59078-970, Natal, RN, Brasil}
\affiliation{$^2$International Institute of Physics, Universidade Federal do Rio Grande do Norte,
Campus Universitario, Lagoa Nova, Natal-RN 59078-970, Brazil}
\affiliation{$^3$Millennium Institute for Subatomic Physics at the High-Energy Frontier (SAPHIR) of ANID, Fern\'andez Concha 700, Santiago, Chile}
\affiliation{$^4$ Departamento de F\'isica, Facultad de Ciencias, Universidad de La Serena,
Avenida Cisternas 1200, La Serena, Chile}
\affiliation{$^5$Universidade Federal de Pernambuco,Centro Acadêmico do Agreste}

\begin{abstract}

Long-lived particles (LLPs) provide an interesting window into physics beyond the Standard Model, offering characteristic signatures at colliders and in cosmology. In this work, we investigate LLPs decays into dark matter. If the lifetime of LLPs are longer than $10^4$~s, the decay products can disrupt the synthesis of light nuclei in the early universe and alter Big Bang Nucleosynthesis (BBN) predictions. If the LLP is much heavier than the dark matter particle, the decay contributes to the number of effective neutrino species, $N_{eff}$. We describe these decays via dimension-six effective operators and outline the parameter space in which such decays obey cosmological bounds stemming from BBN, structure formation, Cosmic Microwave Background, and Baryon Acoustic Oscillation data.
\end{abstract}

\maketitle

\section{Introduction}
\label{sec:intro}

Long-lived particles have gained inscreasing attention due to their interesting signatures at colliders and cosmological observables. However, the lifetime being tested at these laboratories are rather different. While colliders are sensitive to short-lived particles that decay within a few meters of the detector, cosmological probes test very long lived particles with a lifetime longer than a second. Theoretically speaking, very long-lived particles are amenable to a suppression mechanism involving their interactions with Standard Model (SM) particles to justify their long lifetime, conversely to SM particles, which are generally short-lived. This suppression mechanism either occurs in the coupling constant or in the energy scale that drives the interactions. If the natural scale that governs the interaction with the SM particle is very large, say the Planck Scale, the lifetime of the decaying particle is large, as happens in some supersymmetric models, where the neutralino decays into a gravitino plus a photon \cite{Feng:2003uy}. The implication of long-lived particles in cosmology has been considered in several studies \cite{Ellis:1984er,Ellis:1990nb,Kawasaki:1994sc,Asaka:1998ju,Cyburt:2002uv,Bleau:2023fsj}.

A particular, we will explore the implication of LLP decays to the effective number of relativistic species, $N_{eff}$ using an effective field theory (EFT) approach, which captures new physics effects for energies much lower the cut-off scale, $\Lambda$ \cite{Contino,Elis}. These operators are suppressed by inverse powers of $\Lambda$ that govern the decay and, consequently, the cosmological observables. We will assume that such LLPs are embedded in a dark sector with a stable dark matter candidate. Therefore, it is reasonable to assume these dark states come in pairs and for this reason, dimension-six effective operators describe more realistic interactions. For instance, the LLP may decay into the lightest odd particle (dark matter) plus a lepton, photon, or quark or even decay into unstable dark states that later decay into Standard Model particles \cite{Feng:2003uy,Feng:2004zu,Cembranos:2005us,Kelso:2013nwa,Garny:2018ali}. As far as Big Bang Nucleosynthesis (BBN) and the Cosmic Microwave Background (CMB) are concerned, final state electrons and photons will induce similar effects. Accounting for hadronic decays is highly non-trivial due to the complexity of dealing with hadronic processes at low energies \cite{Ellis:1984eq,Audouze:1985be,Kawasaki:1986my,Reno:1987qw,Kohri:2001jx,Kawasaki:2004yh,Kawasaki:2004yh,Kawasaki:2008qe,Omar:2025jue,Angel:2025dkw,Freitas:2009jb}.

Thus, for concreteness, we will analyze decays of the type $X_1 \rightarrow X_2 + \gamma$, where $X_1$ is the LLP and $X_2$ is the lightest odd particle, i.e. a dark matter particle that accounts for the overall dark matter density. As dark matter does not interact with photons directly, it is natural to expect a suppression via higher dimension operators (larger than five). In the regime $m_{X_1} \gg m_{X_2}$, we will show that this decay contributes to a raise in $N_{eff}$. In other words, LLP decays can be a source of an extra radiation component in the early universe \cite{Hooper:2011aj,Calabrese2011,Boehm:2012gr,archidiacono2013,giovanetti2024,Calabrese2011}, and consequently alleviate the Hubble tension \cite{Verde2019,DiValentino_2021}. The interesting aspect of this mechanism is the presence of the photon that provides a visible signature of this decay process, contributing to the diffuse photon background and serving as an indirect probe of the dark sector.

One could invoke dimension five effective operators, but they lead to relatively large decay widths and dangerously small lifetimes, requiring an energy scale larger than the Planck scale to reconcile with cosmological bounds \cite{Jesus2025}. We show that dimension-six operators relax this conclusion and, interestingly, allow weak-scale dark matter for energy scales much below $10^{15}$~GeV. That said, we consider two cases: (i) fermion dark matter and (ii) vector dark matter. Later we derive the corresponding decay width, and the relevant cosmological constraints \cite{steigman2002,Cooke2018,Acharya2019,Bucko2023,Simony,Zhao2013,Alcaniz:2019kah}. We will observe that there is a rich interplay between particle physics and cosmology, particularly with $N_{eff}$ \footnote{We would like to dedicate this work to the memory of Raimundo Silva, who was a great collaborator and helped to build a bridge between particle physics and cosmology at UFRN.}.

The paper is structured as follows: {\it section II} introduces the LLP decays into dark matter; {\it section III} discusses the implications for the Hubble constant; {\it section IV} addresses the relevant bounds; {\it section V} describes the effective field theory approach and outlines the viable parameter space before conclusions in {\it section VI}. 

\vspace{2mm}

\section{LLP Decays and $N_{eff}$}\label{sec:relativistic production}

In contexts where the scale of the new physics is unknown or inaccessible directly in particle accelerators, the EFT approach is particularly useful \cite{Penco2020}.  An EFT can be constructed by specifying three fundamental ingredients: its degrees of freedom (which may or may not be imposed by symmetries); the symmetries that constrain the system's dynamics, reflecting the effective action; and the expansion parameter. From this formalism, effects of new physics and possibly heavy-states of the dark sector can be incorporated through the inclusion of higher-dimensional effective operators that connect the visible and dark sectors. 
 
At early times the energy density of radiation in Universe described by $\Lambda$CDM model is composed mostly by radiation, with photon and neutrinos being the dominant relativistic species at the time, i.e.,  $\rho_{\mathrm{rad}}=\rho_{\gamma}+\rho_{1\nu}N_{\nu}$, in which $\rho_{\gamma}$ represents the energy density of the photons, while $\rho_{1\nu}$ denotes the energy density of a single standard model neutrino species, where $N_{\nu}$ stands for neutrino species number.

Therefore, the energy density is the sum of background photons and neutrinos \cite{dodelson2020, Planck2018} and any extra component, beyond the $\Lambda CDM$ as,
\begin{equation}
    \rho_{\mathrm{rad}} = \rho_{\gamma} + N_\nu  \rho_{1\nu} + \rho_{\mathrm{extra}}.
\end{equation}

This extra source of radiation is known in the literature as dark radiation. We can write this dark radiation in terms of the radiation caused by one neutrino species \cite{Alcaniz2022} ,
\begin{equation}
    \rho_{\mathrm{extra}}=\rho_{1\nu}\Delta N_{eff}, \label{rho extra}
\end{equation}where $N_{eff}$ is the effective number of relativistic species. $\Delta N_{eff}= N_{eff}-N_\nu$. This term {\it effective} is appropriate because $N_{eff}$ is not an integer number. In the $\Lambda CDM$, $N_{eff}$ is sensitive to the time of neutrino decoupling, thus it has a temperature dependence. That said, the total radiation is found to be,
\begin{equation}
    \rho_{\mathrm{rad}}=\rho_{\gamma}+N_{eff} \rho_{1\nu}.
\end{equation}
Studies related to the Cosmic Microwave Background constrain $N_{eff}$ in terms of the cosmological parameters such as the Hubble constant \cite{Calabrese:2012vf,Feeney:2013wp}. 
Adopting the matter-radiation equality as a reference for the computation of $\Delta N_{\mathrm{eff}}$, from Eq. \eqref{rho extra} we can define,
\begin{equation}
   \Delta  N_{\mathrm{eff}} = \frac{\rho_{\text{extra}}(t_\mathrm{eq})}{\rho_{1\nu}(t_\mathrm{eq})} \cdot
    \label{eq:deltaN}
\end{equation}

In summary, a positive contribution to $N_{eff}$ will automatically increase $\rho_{rad}$ and, as a result, the Hubble constant \cite{Planck2018} extracted from CMB measurements. In more general terms, $N_{eff}$ does not represent a neutrino species. It reflects the number of relativistic degrees of freedom at a given temperature. In the scenario considered here, the contribution to $N_\mathrm{eff}$ arises from a relativistic population of the particle $X_2$ produced from the decay of the LLP. As the universe expands, the energy of the $X_2$ particles are redshifted and diminished over time, up to the point of contributing to $N_{eff}$ until the epoch of matter-radiation equality, mimicking the effect of an extra neutrino species and thereby alleviating the Hubble tension \cite{daCosta:2023mow}.

We describe below how this LLP decay can contribute to $N_{eff}$. In the rest frame of the LLP particle, the four-momenta of the particles read,
\begin{align}
    &p_{X_1} = (m_{X_1}, \bm{0}),\nonumber\\
    &p_{X_2} = (E_{X_2}(\bm{p}), \bm{p}),\\
    &p_{\gamma} = (|\bm{p}|, -\bm{p})\nonumber,
\end{align}
where $E_{X_2}$ and $\bm{p}$ stand for the energy and the trimomentum associated with the dark matter particle $X_2$, respectively. 
Immediately after the decay, the four-momenta conservation gives us,
\begin{gather}
    \left|\bm{p}_{X_2}(\tau) \right| = \left|\bm{p} \right| = \frac{1}{2} m_{X_1} \left[ 1 - \left(\frac{m_{X_2}}{m_{X_1}} \right)^2 \right], \label{eq:momentum_chi}\\
    E_{X_2}(\tau) = m_{X_2} \left( \frac{m_{X_1}}{2m_{X_2}} + \frac{m_{X_2}}{2m_{X_1}} \right), \label{eq:energy_in_tau}
\end{gather}
in which $\tau$ is the lifetime of the LLP, $X_1$. In the case where $m_{X_1}\gg m_{X_2}$, the initial energy will be equally distributed between the photon and the dark matter, with both contributing to the radiation content. In addition, it is useful to define the Lorentz factor,
\begin{equation}
    \gamma_{X_2}(\tau)=\frac{m_{X_1}}{2m_{X_2}} + \frac{m_{X_2}}{2m_{X_1}}, \label{gamma}
\end{equation}
in such manner that $E_{X_2} (\tau)=m_{X_2}\gamma_{X_2} (\tau)$.

The expansion of the universe redshifts the momentum of the particle, becoming inversely proportional to the scale factor  $\bm{p}^2_{X_2} \propto 1/a^2$ \cite{hobson2006}. As a consequence,
\begin{equation}
    \begin{split}
        &E^2_{X_2} - m^2_{X_2} = \bm{p}^2_{X_2} \propto \frac{1}{a^2}\\
        &\Rightarrow  \left( E^2_{X_2}(t) - m^2_{X_2} \right)a^2(t) = \left( E^2_{X_2}(\tau) - m^2_{X_2} \right)a^2(\tau)\\
        &\Rightarrow E_{X_2}(t) = m_{X_2}\left[ 1 + \left( \frac{a(\tau)}{a(t)} \right)^2 \left( \gamma^2_{X_2}(\tau) - 1\right)\right]^{1/2}.
    \end{split}
\end{equation}
Computing the ratio $a(\tau)/a(t)$ in the last equality and adopting Eq. \eqref{gamma}, we can define a Lorentz factor $\gamma_{X_2}(t)$ as, 
\begin{equation}
    \gamma_{X_2}(t) = \left\{1+\frac{ (m^2_{X_1} - m^2_{X_2})^2 }{4m^2_{X_2}m^2_{X_1}} \left( \frac{\tau}{t} \right) \right\}^{1/2},
\end{equation}
which is the boost factor acquired by $X_2$ due the decay of the LLP. In the limit $m_{X_1} \gg m_{X_2}$ it reduces to,
\begin{equation}
    \gamma_{X_2}(t_{eq}) -1 \approx \frac{m_{X_1}}{2m_{X_2}} \sqrt{\frac{\tau}{t_{eq}}}.
    \label{eqgamma}
\end{equation}

The total energy of the dark matter particles ought to include those produced from the LLP decay so that
\cite{Tegmark:2000pi,Kopp_2018},

\begin{equation}
    E_{\mathrm{DM}} = N_{\mathrm{HDM}}m_{X_2}(\gamma_{X_2} -1) + N_{\mathrm{CDM}}m_{X_2},
\end{equation}where $N_{\mathrm{HDM}}$ is the number of dark matter particles produced via the LLP decay. $N_{\mathrm{CDM}}$ accounts for the number of dark matter particles produced via a thermal mechanism. The ratio between the energy densities of the two aforementioned components is given by,
\begin{equation}
    \frac{\rho_{\mathrm{HDM}}}{\rho_{\mathrm{CDM}}} = \frac{n_{\mathrm{HDM}}m_{X_2}\left( \gamma_{X_2} -1 \right)}{n_{\mathrm{CDM}}m_{X_2}} \equiv f\left( \gamma_{X_2} -1 \right),
    \label{Eq.ratio}
\end{equation}where $n_{\mathrm{HDM}}$ and $n_{\mathrm{CDM}}$ are the number densities and $f$ the ratio between these two quantities. In other words, $f$ determines the fraction of dark matter particles produced via the LLP decay.

Therefore, using Eq. \eqref{Eq.ratio} together with the definition of $\Delta N_{\mathrm{eff}}$ in Eq. \eqref{eq:deltaN}, we obtain that the extra relativistic degrees of freedom, given by this formalism, can be computed as,
\begin{equation}
    \Delta N_{\mathrm{eff}} = \frac{\rho_{\mathrm{CDM}}f(\gamma_{X_2} - 1)}{\rho_{1\nu}}\cdot \label{eq:deltaN2}
\end{equation}

At matter-radiation equality, the ratio between the energy density of cold dark matter and that of one neutrino is $\rho_{1\nu}/\rho_{CDM}=0.16$, which precisely mimics the effect of one neutrino species in the Hubble rate at later times \cite{Hooper:2011aj}. Using this information and Eq.\eqref{eq:deltaN2} we find,
\begin{equation}
    \Delta N_{eff} = \frac{f\left( \gamma_{X_2} -1 \right)}{0.16}\cdot
    \label{eq:delta16}
\end{equation}

Substituting Eq.\eqref{eqgamma} into Eq.\eqref{eq:delta16} we get,

\begin{equation}
        \Delta N_{eff} \approx 2.5 \times 10^{-3}\sqrt{\frac{\tau}{10^{6} \mathrm{s}}} \times f\frac{m_{X_1}}{m_{X_2}},\label{eq:Neff}
    \end{equation}

Therefore, particle physics and cosmology are intertwined. A particle physics process has a direct impact on $N_{eff}$ and consequently on the Hubble constant extracted from CMB measurements. Our goal in this work is set limits on LLP decays to dark matter.

\section{Constraints}\label{sec:constraints}

\subsection{CMB and BAO Surveys}

Many works have explored the positive correlation between $N_{eff}$ and $H_0$ to solve the $H_0$ problem. Estimates of the Hubble constant $H_0$ inferred from early-Universe observations such as those stemming from the Cosmic Microwave Background \cite{Planck:2018vyg} are systematically lower than those obtained from independent low-redshift probes \cite{Freedman:2023jcz,Riess:2021jrx,Dainotti:2025qxz,Murakami:2023xuy,Gonzalez:2024qjs}. This growing mismatch, now supported by multiple observational techniques, suggests a possible limitation of the standard cosmological model rather than a mere statistical fluctuation \cite{Blanchard:2022xkk,Perivolaropoulos:2021jda}. In particular, a study using Cepheid Observations from the James Webb Telescope obtained $H_0=(73.49\pm 0.93) km s^{-1}Mpc^{-1}$ \cite{Li:2025lfp}. A community report found $H_0= 73.50 \pm 0.81 km s^{-1}Mp^{-1}$ \cite{H0DN:2025lyy}, which translates into a $7.1\sigma$ discrepancy. 

It is known that the Hubble rate at the last-scattering surface can be parametrized as \cite{Riess:2022oxy},
\begin{equation}
H_{ls}=100km s^{-1} Mpc^{-1} \omega_r^{1/2} (1+ z_{ls})\sqrt{1 + \frac{w_m}{w_r}\frac{1}{1+z_{ls}}}
\end{equation} where $z_{ls}=1080$, $\omega_m=\Omega_m h^2$, and

\begin{equation}
  w_r=\left[1+\frac{7}{8}N_{eff}\left(\frac{4}{11}\right)^{4/3}\right]w_\gamma  
\end{equation}with $w_\gamma=2.45\times 10^{-5}$.

 Hence, it is clear that the $H_0$ value inferred from CMB is indeed positively correlated with $N_{eff}$. However, one cannot arbitrarily increase $N_{eff}$ simply focusing on its impact on $H_0$. An increase in $N_{eff}$ can also significantly impact the sound horizon which disagrees with BAO measurements. Therefore, having a large $N_{eff}>0.3$ as solution to $H_0$ is deemed implausible \cite{DiValentino:2021izs}. Moreover, BAO measurements depend on the sound horizon at the drag epoch, which is the distance that sound can travel between the Big Bang and the drag epoch. It quantifies the time when the baryons decoupled from photons. A combination of BAO and CMB data imposed $N_{eff}=3.1\pm 0.17$ \cite{DESI:2024mwx}. A recent study of BBN data taking into account dark radiation has imposed looser bound, namely $\Delta N_{eff}<0.74$ at 95\% C.L. \cite{Ganguly:2025mdi}. In this range, $N_{eff}$ can still help alleviate the Hubble tension \cite{CosmoVerseNetwork:2025alb}, but not solve it. Apparently, the $H_0$ problem requires much more complex solutions than a simple connection to $N_{eff}$ \cite{DiValentino:2021izs,CosmoVerseNetwork:2025alb}. That said, for our reasoning, we will impose the CMB-BAO limit $\Delta N_{eff}<0.3$ \cite{DESI:2024mwx} from now on. The precise value will not change our qualitative conclusions, as we describe below.

\subsection{BBN constraints}
Big Bang nucleosynthesis (BBN) constitutes a cornerstone of early‑universe cosmology, and its successful agreement with light‑element abundances tightly constrains any late injection of energy around the BBN epoch. In particular, the decay channel $X_1 \rightarrow X_2 + \gamma$ can trigger electromagnetic cascades \cite{Feng:2003uy,Cyburt:2002uv}. This extra electromagnetic component may modify the primordial abundances of light nuclei such as helium and deuterium. Before presenting our final constraints, we briefly summarize how these bounds are obtained. Prior to the decay of $X_1$, the universe already contains a thermal photon background; after the decay, the photons observed in the CMB comprise both of this original background and the contribution from the decay‑induced cascade. Accordingly, we express the mean CMB photon energy as,
\begin{equation}
    E^{CMB}_{\gamma} = E^{BG}_{\gamma}\left( \frac{n^{BG}_{\gamma}}{n^{CMB}_{\gamma}} \right) + E_{\gamma}\left( \frac{n_{\gamma}}{n^{CMB}_{\gamma}} \right),
\end{equation}where $E^{BG}_{\gamma}$ denotes the mean energy of the background photons, $E_{\gamma}$ is the mean energy of the photons produced in the decay, $n^{BG}_{\gamma}$ is the background photon number density, $n^{CMB}_{\gamma}$ is the total CMB photon number density, and $n_\gamma$ represents the number density of photons generated in our scenario. Knowing that the decay $X_1 \rightarrow X_2 + \gamma$ implies that $n_{X_1} = n_{X_2} = n_{\gamma}$, where $n_\gamma$ is the number density of photons added to the universe due to the decay, we obtain the overall electromagnetic energy injected by the LLP,
\begin{equation}
    \zeta_{EM} \equiv E_{\gamma}Y_{\gamma}=E_\gamma Y_{X_1} ,
    \label{eq:zeta_definition}
\end{equation}where $Y_{\gamma} = n_{\gamma}/n^{CMB}_{\gamma}$, and $Y_{X_1} = n_{CDM}\times f/ n^{CMB}_{\gamma}$.

Eq. \eqref{eq:zeta_definition} quantifies the electromagnetic energy injected by the decay. Using the definition of critical density $(\rho_c \equiv 3H/(8\pi G))$, the definition of density parameter $(\Omega \equiv \rho/\rho_c)$, the cold particle energy density $(\rho= nm)$, and the time evolution of number density of CMB photons $( n^{CMB}_\gamma = n^{CMB}_{\gamma,0}/a^3 )$ \cite{ref:hobson2006GR}, we can rewrite $Y_{X_1}$ as,

\begin{equation}
    Y_{X_1} = \frac{f}{m_{X_1} n^{CMB}_{\gamma,0}} \times \Omega_{CDM,0} \rho_{c,0},
\end{equation}where $ \Omega_{CDM}a^3 \rho_c = \Omega_{CDM,0} \rho_{c,0}$ \cite{ref:hobson2006GR}.

Plugging in the known quantities, $\rho_{c,0} \simeq 1.05 \times 10^{-5}h^2 GeV/cm^3$, $n^{CMB}_{\gamma,0} = 411 cm^{-3}$  and $\Omega_{CDM,0}h^2 = 0.12$ we find, 
\begin{equation}
    Y_{{X_1}} = 3.01 \times 10^{-9} \left( \frac{GeV}{m_{X_1}} \right) \times f.
\end{equation}

Assuming, $m_{X_1} \gg m_{X_2}$, the injected energy into the plasma per decay is $E_\gamma = m_{X_1}/2$. Consequently, the electromagnetic radiation is given by, 
\begin{equation}
    \zeta_{EM} = E_\gamma Y_{X_1} = 1.5 \times 10^{-9} GeV \times \left( f \frac{m_{X_1}}{m_{X_2}} \right).
    \label{eqzeta}
\end{equation}

Thus, the total electromagnetic energy injected into the process, as determined by Eq.\eqref{eq:zeta_definition}, is tied to the properties of the dark sector. Therefore, we can constrain the properties of the dark sector using BBN observations. 

In Figure~\ref{fig:CMBBBN} we exhibit the BBN limits in terms of the ratio $f m_{X_1}/m_{X_2}$ and the lifetime of the decay, assuming prompt decay. The shaded regions correspond to parameter space ruled out due to the over-destruction of helium‑4, lithium‑7, and deuterium, or trigger nuclear reactions that overproduce deuterium in conflict with astronomical abundance measurements \cite{Holtmann:1998gd,Kawasaki:2004yh,Kawasaki:2004qu,Kohri:2006cn,Kawasaki:2007xb,Kawasaki:2008qe,Jittoh:2011ni}. At very early times (before $10^4 s$) injected photon thermalize with background photons doing no harm to the abundances of light elements. At later times, thermalization becomes inefficient, and for this reason, the constraints become stronger with the lifetime. The green region represents the parameter space that produces $\Delta N_{eff}=0.1-0.3$.

\subsection{Spectral Distortions}

\begin{figure}
    \centering
    \includegraphics[width=\linewidth]{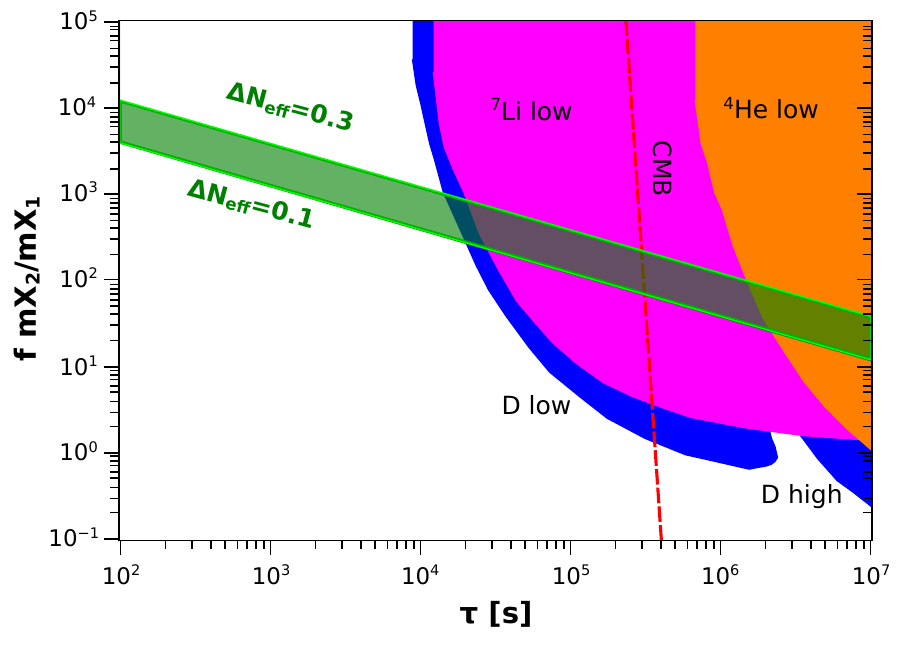}
    \caption{BBN bounds based on light element abundances and CMB constraint stemming from spectral distortion of the CMB. See text for details. The green region delimits the parameter space that yields $\Delta N_{eff}=0.1-0.3$. Using the positive correlation between $\Delta N_{eff}$ and $H_0$ one could raise $H_0$ and alleviate the Hubble tension.}
    \label{fig:CMBBBN}
\end{figure}

The injection of electromagnetic energy may also distort the frequency dependence of the CMB spectrum. Double Compton scattering ($\gamma e^-\rightarrow \gamma\gamma e^-$), and bremsstrahlung ($e^-X \rightarrow e^-X\gamma$) are not very efficient for $\tau > 10^4$s. As a result the CMB spectrum relaxes to a Bose-Einstein distribution function with a chemical potential different from zero with,
\begin{equation}
 \mu = 8 \times 10^{-4}\left(\frac{\tau^4}{10^6s} \right)^{1/2}\left(\frac{\zeta_{EM}}{10^{-9}GeV} \right)e^{\left(-\tau_{dc}/t\right)^{5/4}},
\end{equation}where $\tau_{dc}$ is a constant that depends on the baryonic density\cite{Feng:2003uy}. Using Eq.\eqref{eqzeta} we can plot the CMB bound in the plane $f m_{X_1}/m_{X_2} \times \tau$ as shown in Figure ~\ref{fig:CMBBBN}. In the region of interest, the CMB limit is less competitive than the BBN one.

\subsection{Structure Formation}

The median velocity of the decay product at a given time $t$ for the two-body decay is,
\begin{equation}
v_{1,2,{\rm med}}(t) \sim {a_{\rm dec}/a(t)p \over \sqrt{(a_{\rm dec}/a(t)p)^2+m_{1,2}^2}}.
\end{equation}

The important quantity regarding structure formation is the free-streaming distance, which can be written as,
\begin{equation}
k_{\rm fs}(a)={\sqrt{\rho a^2/2 M^2_{\rm P}} \over v_{\rm med}(a)} = \sqrt{{3\over2} }{aH(a) \over v_{\rm med}(a)}, 
\end{equation} where for $k > k_{\rm fs}$, the growth of structure is suppressed. Hot dark matter (HDM) particles have large free-streaming. The abundance of hot dark matter
is often quoted in terms of the massive neutrinos abundance
\cite{Reid_2010,Zhao_2013}. The best cosmological probe of constraining massive neutrinos acting as HDM is the galaxy power spectrum. The latest result from the DESI collaboration imposes the sum of the neutrino masses to be $\Sigma \, m_{\nu} < 0.113$~eV~\cite{DESI:2024mwx}, which translates into $\Omega_{\nu}h^2 < 0.001$, and consequently $f=\Omega_{HDM}/\Omega_{CDM}<0.009$. Roughly speaking, a mechanism that produces a population of HDM is limited to account for at most $0.9\%$ of the overall dark matter abundance. For this reason, we will set $f=0.009$ hereafter.

\subsection{Entropy Injection}
\label{secentropy}

In the radiation-dominated era the difference in entropy between two initial times, before the decay, and a final time, after the decay, is found to be,

\begin{equation}
   \frac{S_f}{S_i}=\frac{g_{\star s}^f T_f a_f^3}{g_{\star s}^i T_i a_i^3},
\end{equation}where $T$ is the photon temperature, while $g_{*s}$ is defined in terms of the bosonic ($g_b$) and fermion ($g_f$) degrees of freedom as. As the temperature is related to the photon density, which in turn depends on $g_{\ast
}$, after the LLP decay, the change in entropy ratio becomes \cite{Alcaniz:2022oow},

\begin{equation}
   \frac{S_f-S_i}{S_i}=\frac{\Delta S}{S_i}=\frac{g_{\ast s}^f}{g_{\ast s}^i}\left( \frac{g_{\ast }^i}{g_{\ast }^f} \right)^{3/4}-1 \simeq 0.05 \Delta N_{eff}.
\end{equation}

Having in mind the existing limits on $\Delta N{eff}$, we conclude that the change in entropy is small, in agreement with CMB observations and BBN. 

\section{Dimension-6 Effective Lagrangians}\label{sec:Leff}

The results developed so far are quite general and model-independent, applying to all effective Lagrangians that describe a LLP decaying into relativistic DM and a photon. The specific details of the model arise when computing the decay width and inserting it into Eq.~\eqref{eq:Neff} in the form of the lifetime. In this sense, it is possible to obtain a relation between the cutoff energy scale $\Lambda$, the extra effective number of neutrino species $\Delta N_{\mathrm{eff}}$, and the masses of the particles involved in the decay. We do this exercise for: a) fermion dark matter and b) for vector dark matter. \\

\textbf{Case A: Fermion Dark Matter}
A fermion LLP may decay into a fermion dark matter particle and a photon via the Lagrangian
\begin{figure}[htb!]
    \centering
        \begin{tikzpicture}
            \begin{feynman}[large]
                \vertex[blob] (a) {};
                \vertex [left = of a] (b) {$X_1$};
                \vertex [above right = of a] (c) {$X_2$};
                \vertex [below right = of a] (d) {$\gamma$};
                \vertex [right = of a] (e);
                \diagram{
                    (b) --[fermion] (a) --[fermion] (c);
                    (a) --[boson] (d);
                };
            \end{feynman}
        \end{tikzpicture}
        \label{fig:diagram6_FFg}
    \caption{Diagrammatic representation of the decay $X_1 \rightarrow X_2 + \gamma$. In order to have relativistic production of dark matter, it is assumed that $m_{X_1}\gg m_{X_2}$.}
    \label{fig:diagrams_neutrinos}
\end{figure}
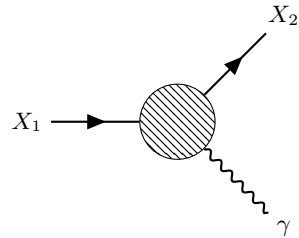

\begin{align}\label{ffgamma}
\mathcal{L}_{\mathrm{eff}} &= \frac{1}{\Lambda^{2}} \left\{ i\,\bar{X}_{1}\,\gamma^{\mu} \,\partial^{\nu} X_{2} \, F_{\mu\nu} + \text{h.c.} \right\},
\end{align}as illustrated in Figure~\ref{fig:diagrams_neutrinos}.

We remind that $X_1$ and $X_2$ represent the LLP and dark matter field, respectively, and $F^{\mu\nu}$ the electromagnetic field strength. This dimension-6 operator couples the fermion bilinear $\bar{X}_{1}\gamma^{\mu}\partial^{\nu}X_{2}$ to $F_{\mu\nu}$, giving rise to the two-body decay $X_{1}\rightarrow X_{2}+\gamma$. After the LLP decay, the kick transferred to the dark matter particle will contribute to $\Delta N_{eff}$ as discussed in Section \ref{sec:relativistic production}. A straightforward computation of the decay width from Lagrangian Eq.\eqref{ffgamma} gives,
\begin{align}
\Gamma &= \frac{m_{X_{1}}^{5}}{32\pi \Lambda^{4}}  
\left[ 1 - \left(\frac{m_{X_{2}}}{m_{X_{1}}}\right)^2 \right]^{3}
\left[ 1 - \left(\frac{m_{X_{2}}}{m_{X_{1}}}\right) \right]^{2} \approx \frac{m_{X_{1}}^{5}}{32\pi\Lambda^{4}}.
\end{align}

Inserting this relation into Eq.~\eqref{eq:Neff}, we relate the LLP and dark matter masses for a fixed cutoff scale. Accordingly, 
\begin{align}
m_{X_{1}} &= \left( \frac{32\pi\Lambda^{4}}{d\, m_{X_{2}}^{2}} \right)^{1/3}, \label{eq:mX1fLLP}
\end{align}
with $d$ being an auxiliary constant depending on $\Delta N_{\mathrm{eff}}$ as follows,
\begin{align}
d &= \frac{\Delta N_{\mathrm{eff}}^{2} \times 10^{12}\,\mathrm{s}}{25 f^{2} \,\hbar \; (\mathrm{GeV}\cdot\mathrm{s})}.
\label{eqd}
\end{align}

Using Eq.\eqref{eq:mX1fLLP} and Eq.\eqref{eqd} we relate the relevant quantities $\Lambda$, $m_{X_1}$ and $m_{X_2}$ to $\Delta N_{eff}$.

\begin{figure}[H]
    \centering
    \subfigure[]{
    \includegraphics[width=0.95\columnwidth]{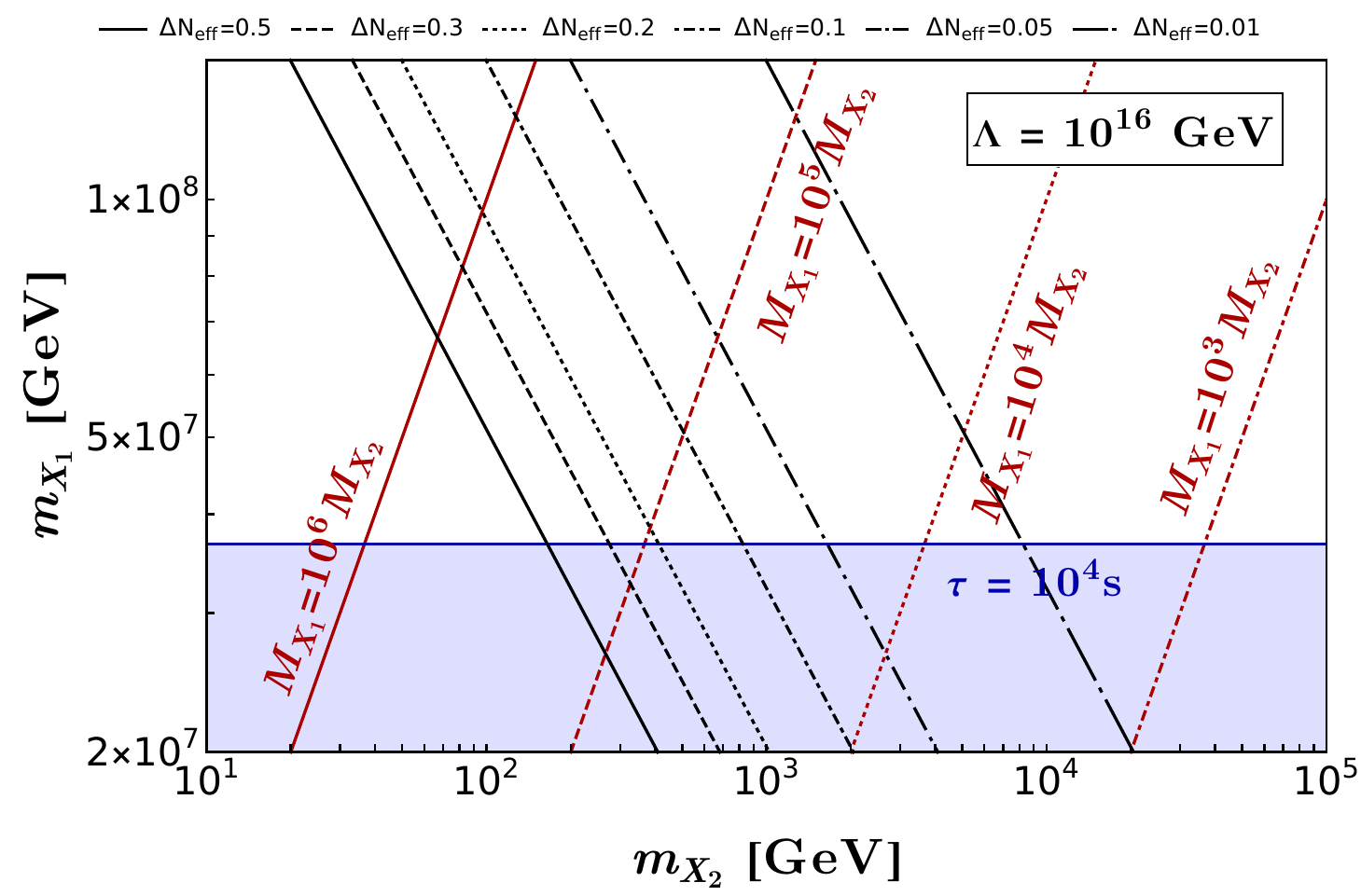}
    \label{fig3:L10to16}
}
    \subfigure[]{
    \includegraphics[width=0.95\columnwidth]{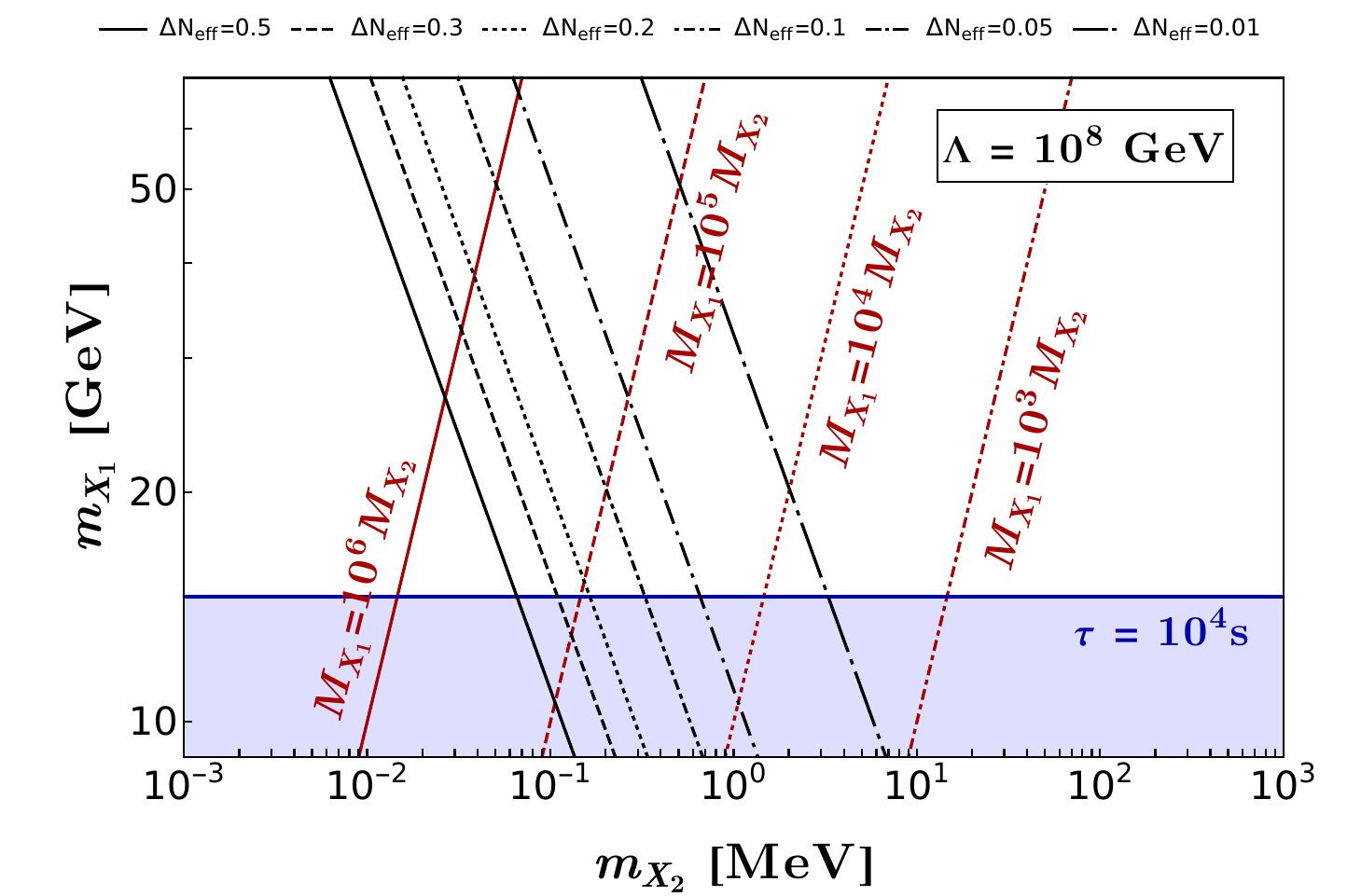}
    \label{fig3:L10to8}}
    \caption{Parameter space for the decay \eqref{ffgamma}.Allowed regions in the long-lived particle ($m_{X_{1}}$) and dark matter ($m_{X_{2}}$) parameter space for different values of the energy scale $\Lambda$. The colored shaded area represents the BBN constrains $\tau \leq 10^4\,s$ \cite{Alcaniz2022}. The red lines indicates different values of the mother-to-daugther mass ratio.}
    \label{figure3}
\end{figure}

We learned from Figure~\ref{fig:CMBBBN} that a lifetime longer than $10^4$~s is excluded by BBN, having in mind the desired $f\, m_{X_2}/m_{X_1}$ ratio to yield $\Delta N_{eff}=0.1-0.3$. Lower values of $\Delta N_{eff}$ can be obtained by decreasing the $f\, m_{X_2}/m_{X_1}$ ratio. In Figure \ref{fig3:L10to16} we delimit the parameter space  that yields a lifetime $\tau \leq 10^4$~s and $\Delta N_{eff}=0.01-0.5$ for $\Lambda=10^{16}$~GeV. We repeated this exercise for $\Lambda=10^8$~GeV in Figure~\ref{fig3:L10to8}. We remind the reader that we have assumed $f=0.009$ throughout, as discussed previously.

In Figure~\ref{fig3:L10to16}, we cover dark matter masses going from $10$~GeV up to $10^5$~GeV. The shaded region is excluded by BBN. That said, we conclude that a large region of parameter space is free from cosmological bounds, but only a small one can contribute to $\Delta N_{eff}=0.1-0.3$ and potentially raise the $H_0$ value. Interestingly, imposing $\Delta N_{eff}<0.1$, we can set a lower mass limit on the dark matter that yields,  namely $m_{X_2}> 1$~TeV. Once we lower the effective energy scale down to $\Lambda=10^8$~GeV, we allow lower masses, as shown in Figure~\ref{fig3:L10to8}. In Figure~\ref{fig3:L10to8}, dark matter masses should lie in the MeV region to induce a 
$\Delta N_{eff}<0.1$. Having dark matter masses below the TeV scale is an advantage of using effective operators of higher dimension compared to dimension-five effective operators \cite{Jesus2023}.

\vspace{1cm}
\textbf{Case B: Vector Dark Matter}

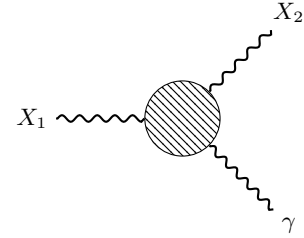
\begin{figure}[htb!]
    \centering
        \begin{tikzpicture}
            \begin{feynman}[large]
                \vertex[blob] (a) {};
                \vertex [left = of a] (b) {$X_1$};
                \vertex [above right = of a] (c) {$X_2$};
                \vertex [below right = of a] (d) {$\gamma$};
                \vertex [right = of a] (e);
                \diagram{
                    (b) --[boson] (a) --[boson] (c);
                    (a) --[boson] (d);
                };
            \end{feynman}
        \end{tikzpicture}
        \label{fig:diagram6_FVN}
    \caption{Diagram representing of the decay $X_1 \rightarrow X_2 + \gamma$ decay, where $X_1$ and $X_2$ are both vector fields.}
    \label{fig:diagrams_neutrinos_2}
\end{figure}

If dark matter is a vector field, its production via LLP decay (see Figure~\ref{fig:diagram6_FVN}) can be described by the Lagrangian form,

\begin{align}\label{vvgamma}
\mathcal{L}_{eff} &= \frac{1}{\Lambda^2} \, G^{\mu}{}_{\nu} \, X^{\nu\alpha} \, F_{\mu\alpha} \, ,
\end{align}

\begin{figure}[H]
    \centering
    \subfigure[]{
    \includegraphics[width=0.95\columnwidth]{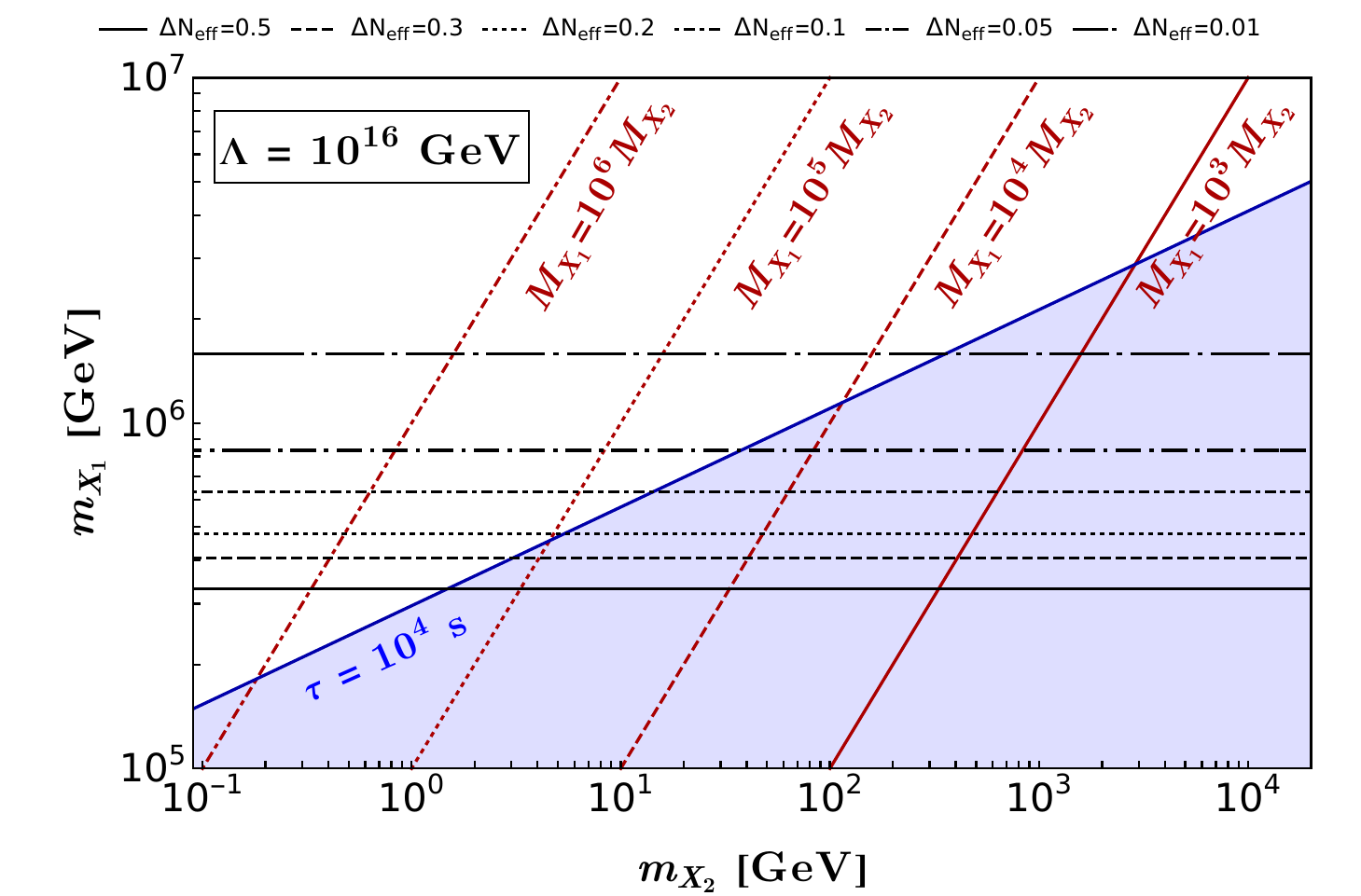}
    \label{fig:fDM_L10to16_2}}
    \subfigure[]{
    \includegraphics[width=0.95\columnwidth]{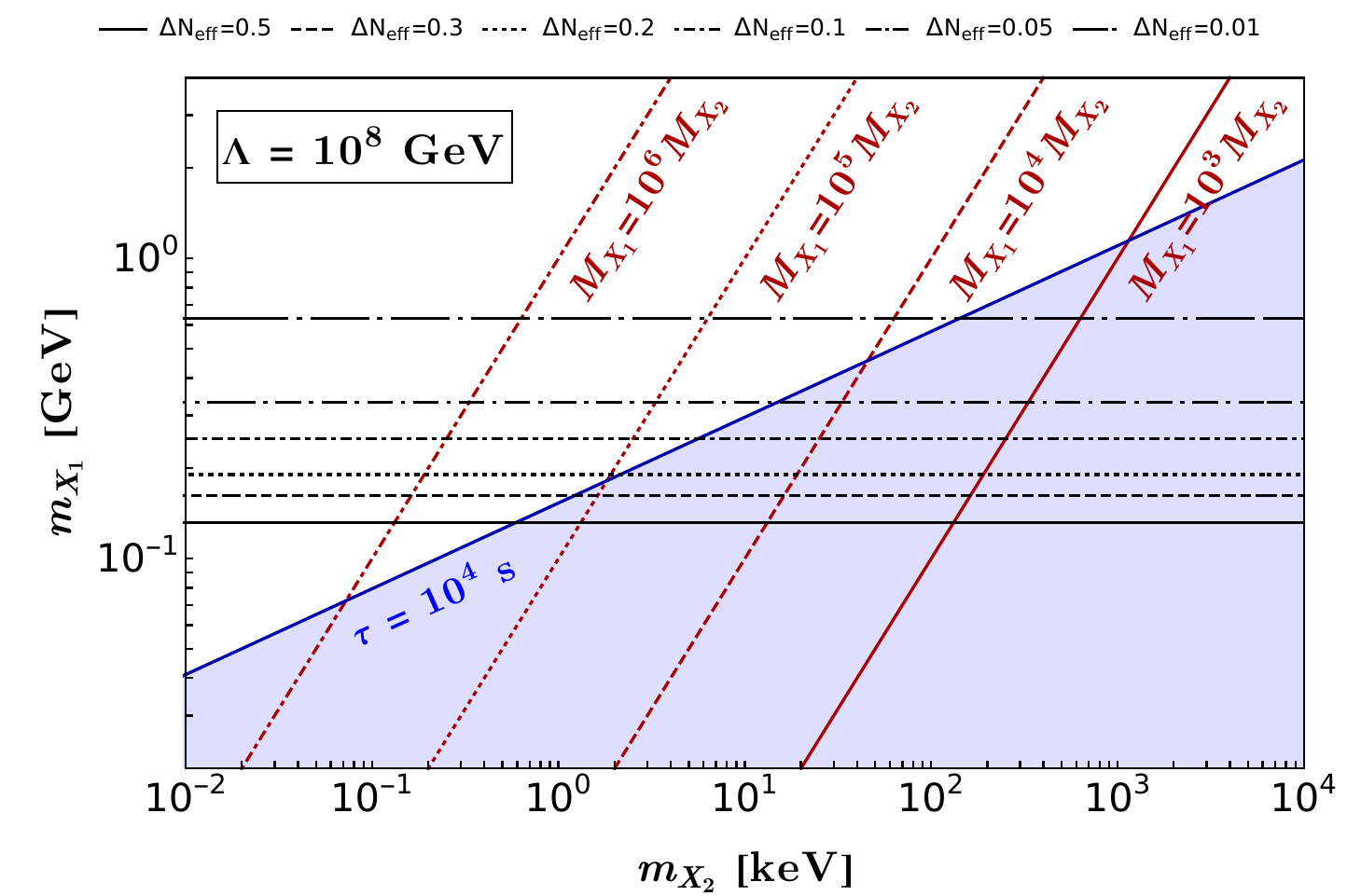}
    \label{fig:fDM_L10to8_2}}
    \caption{Parameter space for the decay \eqref{vvgamma}. Again, The colored shaded area represents the BBN constrains $\tau \leq 10^4\,s$, whereas the red lines delineates different values of the ratio $m_{X_1}/m_{X_2}$.}
    \label{fig:vectorplots}
\end{figure}

From Eq.\eqref{vvgamma} we find the decay width,
\begin{align}
\Gamma &= \frac{1}{96\pi\Lambda^{4}} \, \frac{m_{X_{1}}^{7}}{m_{X_{2}}^{2}} 
\left( 1 - \frac{m_{X_{2}}^{2}}{m_{X_{1}}^{2}} \right)^{3} \nonumber\\ &  \times
\left[ 1 - 4\left( \frac{m_{X_{2}}}{m_{X_{1}}} \right)^{4} - \left( \frac{m_{X_{2}}}{m_{X_{1}}} \right)^{6} \right] \nonumber\\
&\simeq \frac{m_{X_{1}}^{7}}{96\pi\Lambda^{4}m_{X_{2}}^{2}} \, .
\end{align}
Similarly to the  preview process, we get a relation with respect to the energy scale as follows,
\begin{align}
m_{X_{1}} &= \left( \frac{96\pi\Lambda^{4}}{d} \right)^{1/5},
\end{align}where $d$ is given in Eq.\eqref{eqd} with $m_{X_1}$ not depending explicitly on $m_{X_2}$, implicating in straight lines in the $m_{X_1}$ versus $m_{X_2}$ plane. In Figure~\ref{fig:fDM_L10to16_2}, we plot the region of parameter space in the $m_{X_1}-m_{X_2}$ plane which reproduces $\Delta N_{eff}>0$ assuming $\Lambda =10^{16}$~GeV.

As aforementioned, the main distinction between the fermion dark matter and the vector dark matter is the relation between $m_{X_1}$ and $\Delta N_{eff}$ ruled by Eq.\eqref{eqd}. In Figure\ref{fig:fDM_L10to16_2}, we take $\Lambda=10^{16}$~GeV. We conclude that the dark matter mass should be larger than $\sim 10$~GeV to induce $\Delta N_{eff}<0.1$. Whereas for  $\Lambda=10^{8}$~GeV (see Figure~\ref{fig:fDM_L10to8_2}), only when $m_{X_2} > 10$~keV we can reproduce $\Delta N_{eff} <0.1$. We highlight that, the entire parameter space with $\tau< 10^4$~s, which yields $\Delta N_{eff} <0.3$ \cite{DESI:2024mwx} is consistent with BBN, CMB and BAO data. There are clearly two ways clear outcomes: (i) once we enforce an upper limit on $\Delta N_{eff}$, say $\Delta N_{eff}<0.1$, we find a lower mass bound on the dark matter mass. The precise value depends on the dark matter nature and effective energy scale $\Lambda$; (ii) if one wants to have LLP decays as a possible source of dark radiation to alleviate the tension between CMB and local measurements of $H_0$ an upper bound on the dark matter mass is found. Considering Figure~\ref{fig:fDM_L10to8_2} for concreteness, and imposing $0.1<\Delta N_{eff} <0.3$, we conclude that the dark matter mass should be smaller than $4$~keV. Thus, observational cosmology  can place constraints on the production mechanism of dark matter particles via its signatures on structures formation, BAO, BBN and particularly $N_{eff}$.


\section{Conclusions}
\label{sec:sec2}
In this work, we investigate how effective dimension-6 operators can map the cosmological impacts of long-lived particles decays into dark matter. We investigated fermion and vector dark matter particles with mass $m_{X_2}$. Structure formation requires that only a fraction of the overall dark matter abundance should be produced via this mechanism. BBN imposes a lifetime shorter than $10^4$~s to avoid conflict with observations of the primordial abundance of Deuterium and $^4$He. A combination of BAO and CMB surveys impose $\Delta N_{eff}<0.3$. That said, we computed the decay widths of the long-lived particles in terms of the relevant parameters and found a clear advantage of considering dimension-six effective operators in comparison with previous studies in the literature that were limited to dimension-five effective theories. Dark matter masses below the TeV scale have become viable. In particular, we found that $\Delta N_{eff}$ can constrain this dark matter production mechanism down to keV masses. Setting $\Lambda=10^8$~GeV, sub-MeV fermion dark matter can yield $\Delta N_{eff}=0.1-0.3$. The larger the dark matter mass, the smaller the effect on $N_{eff}$. Due to the properties of the interactions involved, LLPs decays into vector dark matter favor masses below the keV scale, whereas decays into fermion dark matter prefer masses below 1MeV, taking $\Lambda=10^8$~GeV. 

Two distinct outcomes emerge clearly from our analysis: (i) If we impose an upper bound on ($\Delta N_{\rm eff}$), for instance ($\Delta N_{\rm eff} < 0.1$), we obtain a corresponding lower bound on the dark matter mass. The exact value of this bound depends on the nature of the dark matter candidate and on the effective energy scale ($\Lambda$); (ii) On the other hand, if long-lived particle (LLP) decays are invoked as a possible source of dark radiation with the intention of alleviating the tension between Cosmic Microwave Background measurements and local determinations of the Hubble constant ($H_0$), an upper bound on the dark matter mass arises. Our main results are summarized in Figures~\ref{figure3} and \ref{fig:vectorplots}. In summary, cosmological observations going from structure formation, BBN, BAO to CMB are powerful probes for the production of dark matter resulted from LLPs decays.

\section{Acknowledgements}
The authors thank Saulo Carneiro and Rafael Nunes for comments on the manuscript. RH acknowledges the support from CNPq through grant 308550/2023-4. MP acknowledges the support from CNPq through grant 151811/2024-5. MP also thanks the ICTP-Trieste and Mehrdad Mirbabayi for their hospitality while part of this work was carried out. MF was financed by the Coordenação de Aperfeiçoamento de Pessoal de Nível Superior - Brasil (CAPES) – Finance Code  001. FSQ acknowledges support from Simons Foundation (Award Number:1023171-RC), CNPq 403521/2024-6, 408295/2021-0, 403521/2024-6, 406919/2025-9, 351851/2025-9, the FAPESP Grants 2021/01089-1, 2023/01197-4, ICTP-SAIFR 2021/14335-0, and the ANID-Millennium Science Initiative Program ICN2019\_044. PVS and FSQ were partially funded by FINEP under project 213/2024 and was carried out in part through the IIP cluster {\it bulletcluster}.

\bibliography{ref}

\end{document}